\documentclass[aps,prl,twocolumn,showpacs]{revtex4}
\usepackage{amsmath}
\usepackage{epsfig}
\usepackage{amssymb}
\usepackage{dcolumn}
\usepackage{graphicx}
\usepackage{dcolumn}

\begin{document}
\date{\today}

\title{Dissipative solitons which cannot be trapped}

\author{Rosa Pardo}
\email{rpardo@mat.ucm.es}
\affiliation{Departamento de Matem\'atica Aplicada, 
Facultad de Ciencias Qu\'{\i}micas, 
Universidad Complutense,  Avda. Complutense s/n, 
28040 Madrid, Spain}

\author{V\'{\i}ctor  M. P\'erez-Garc\'{\i}a}
\email{victor.perezgarcia@uclm.es}
\homepage{http://matematicas.uclm.es/nlwaves}
\affiliation{Departamento de Matem\'aticas, Escuela T\'ecnica
Superior de Ingenieros Industriales,
Universidad de Castilla-La Mancha 13071 Ciudad Real, Spain.}

\affiliation{Instituto de Matem\'atica Aplicada a la Ciencia y la Ingenier\'{\i}a (IMACI),
Universidad de Castilla-La Mancha, 13071 Ciudad Real, Spain.}

\begin{abstract}
In this paper we study the behavior of dissipative solitons in systems with high order nonlinear dissipation and show how they cannot survive under the effect of 
trapping potentials both of rigid wall type or asymptotically increasing ones. This provides an 
striking example of a soliton which cannot be trapped and only survives to the action of a weak potential. 
\end{abstract}


\maketitle

%\section{Introduction}

\emph{Introduction.-} Solitons are self-localized nonlinear waves which are sustained by an equilibrium between dispersion and nonlinearity and appear in a great variety of physical contexts  \cite{SoliGen}. 
Solitonlike localized states also arise in dissipative systems driven 
far from thermal equilibrium such as hydrodynamics \cite{1}, granular 
media \cite{2}, gas discharges \cite{3}, nonlinear optics and other fields \cite{4,5,6,7,8,9,10,11,11b,12}.
These structures are referred to as  ``dissipative solitons" and are sustained  because of an interplay between
dispersion, nonlinearity, gain and losses.

 A new family of  multidimensional dissipative (bright) solitons has been recently observed and studied in the context of the propagation of optical beams in media with multiphoton absorption  \cite{Dubietis,Porras,Dubietis2,Polesana}. These solitons have slowly decaying tails and have a nontrivial flux of energy from their tails to the soliton center where energy is absorbed leading to a stable nonlinear object. In some sense they represent a nontrivial extension of the well-known non-diffracting optical beams \cite{ND}.

These dissipative soliton-like structures have been also studied in different contexts such as three-dimensional scenarios \cite{O-waves,Ou} and  should also exist in other physical scenarios ruled by analogous physical phenomena (and  model equations)  \cite{Sulem}. 

In this letter we will describe a highly counterintuitive property of these solitons which is that they \emph{cannot be trapped} by the action of an attractive potential, an affirmation which we will make more precise later. This fact is very striking since attractive potentials use to  enhance the stability of solitons. In fact, it is well known from Quantum Mechanics that attractive potentials alone without any help of nonlinear interactions are able to support stationary solutions. Even unstable nonlinear structures such as the  Townes soliton \cite{Sulem} are stabilized by an external confining potential \cite{Tow1,Tow2}. 

\emph{Mathematical model.-} As a mathematical model of solitons under high order dissipation we will
 use the nonlinear Schr\"odinger equation (NLS) 
\begin{equation}
i\frac{\partial \psi}{\partial t}=-\frac{1}{2}\Delta \psi+ V\psi + g|\psi|^2\psi+i\gamma|\psi|^4\psi,
\label{gp_dimensionless}
\end{equation} 
on $\mathbb{R}^2$, where $g$ and $\gamma>0$ are real parameters (nonlinearity and dissipation coefficients) and $V$ is a real function describing an external potential acting on the system. 
Eq. (\ref{gp_dimensionless}) models the
propagation of optical beams in media with multiphoton absorption \cite{Dubietis,Porras,O-waves}, the mean field dynamics of Bose-Einstein condensates with losses associated to inelastic three-body collisions \cite{Ou,cBEC0,cBEC1,cBEC3,cBEC4,cBEC5,cBEC6,cBEC3b} and other physical phenomena \cite{Sulem}. The mathematical studies on this equation have  focused on the study of
the concentration mechanisms leading to the so called ``super-strong" collapse phenomenon (which however, does not involve the formation of a singularity in the mathematical sense) \cite{Zakharov,dissipation,Sulem}.
 
\emph{Stationary solutions.-} Stationary solutions of Eq. (\ref{gp_dimensionless}) with $\gamma =  0,$ and $V =0$ have been 
discussed extensively  since they correspond to the simplest model of nonlinear optical beams and classical superfluids. When $\gamma \neq 0$  they have been studied in Refs.  \cite{Porras,O-waves,Ou}. 
Let us write
\begin{equation}
\psi(\boldsymbol{r},t)=A(\boldsymbol{r},t)\exp[i\phi(\boldsymbol{r},t)],
\end{equation}
 with $A(\boldsymbol{r},t)>0$. Then Eqs. (\ref{gp_dimensionless}) become
\begin{subequations}
\label{mod-pha}
\begin{eqnarray}
\frac{\partial (A^2)}{\partial t} & = & -\nabla \cdot \left(A^2\nabla \phi\right) + 2 \gamma A^6, \label{modphaa} \\
\frac{\partial \phi}{\partial t} & = & \frac{1}{2} \left[\frac{1}{A}\Delta A - \left(\nabla \phi\right)^2\right] - g A^2 - V.\label{modphab}
\end{eqnarray}
\end{subequations}
Stationary solutions of Eq. (\ref{gp_dimensionless}) satisfy 
% \begin{equation}\label{NL1}
 $\partial_{t} A=0$, and $
 \phi(\boldsymbol{r},t)=\varphi(\boldsymbol{r})-\lambda t$. 

It is obvious that $N \equiv \|\psi\|_2^2 = \int_{\mathbb{R}^2}|\psi|^2 = \int_{\mathbb{R}^2}A^2$, cannot change over stationary solutions. However, let us assume that $N$ is finite for stationary solutions, then using Eq. (\ref{gp_dimensionless}) we get
\begin{equation}
\frac{dN}{dt} = -2\gamma \int_{\mathbb{R}^2} A^6 < 0,
\end{equation} 
 which is a contradiction. This implies that $N$ cannot be finite for nontrivial stationary solutions of Eq. (\ref{gp_dimensionless}). This fact is also  known to be characteristic of the linear non-diffracting optical beams \cite{ND}.

The existence of a nontrivial phase structure $\varphi(\boldsymbol{r})$ is a feature of the case $\gamma \neq 0$  which is not present in the  conservative NLS. 
%Substituting %Eqs. \eqref{NL1} and \eqref{NL2} 
%Eq. \eqref{NL} into  Eqs. (\ref{mod-pha}a)-(\ref{mod-pha}b) 
From Eqs. (\ref{modphaa}) and (\ref{modphab})
\begin{subequations}
\label{mod-pha-st_normalized}
\begin{eqnarray}
 \nabla \left( A^2 \nabla  \varphi\right)  & = & 2\gamma A^6, \label{phase_equation}\\
 - \frac{1}{2} \Delta A + \frac{1}{2} \left(\nabla \varphi \right)^2 A + VA& = & \lambda  A- g A^3 . \label{amplitude_equation}
\end{eqnarray}
\end{subequations} 
In this paper, in order to deal with physically meaningful solutions, we will  
 study solutions of Eqs. (\ref{mod-pha-st_normalized}) which decay asymptotically for large values of $r= \| \boldsymbol{r}\|$, i.e. $A(r), \varphi(r)  \rightarrow 0$ when $r \rightarrow \infty$.%\subsection{Radially symmetric case}

When the amplitude and phase are  radially symmetric functions $A(\boldsymbol{r}) = R(r)$, $\varphi(\boldsymbol{r})  =\Phi(r)$,
Eq. (\ref{phase_equation}) can be integrated by using the divergence theorem and we get 
\begin{equation}\label{Phi}
\Phi'(r) = \frac{2\gamma}{R^2r} \int_0^r \rho R(\rho)^6 \, d\rho.
\end{equation}
Introducing  \eqref{Phi} into \eqref{amplitude_equation}
we can write
\begin{multline}
-R'' - \frac{1}{r}R' + 2 V R =  \\  2\lambda R  - 2g R^3-    \frac{4\gamma^2}{R^3r^2} \left( \int_0^r \rho R(\rho)^6 d\rho \right)^2, \label{r-R:2}
\end{multline}
complemented with the boundary conditions $R(r) \to 0$ as $r\to\infty$ and $%R(0) = R_0, 
R'(0) = 0.$ 
%(the obvious overdetermination implies that not every pair $R_0, \lambda$ leads to solutions of this nonlinear eigenvalue problem). 
 Eq. (\ref{r-R:2}) is a nonlinear eigenvalue problem with a nonlocal term.

Let us look for solutions satisfying that
\begin{equation}
Q = 2\gamma   \int_0^{\infty} r R(\rho)^6 \, dr= \frac{\gamma}{\pi} \|R\|^6_{L^6 (\mathbb R ^2)}  ,
\end{equation}
is finite (mathematically, this means that $R \in L^6 (\mathbb R^2)$). In this case Eq. (\ref{Phi}) allows us to obtain $
\Phi'(r)  \approx  Q /\left[r R^2(r)\right]$,  for $r \gg 1$, and \begin{equation}
-R'' - \frac{1}{r}R' + 2 V R =   2\lambda R  - 2g R^3-    \frac{Q^2}{R^3r^2}. \label{r-R:3}
\end{equation}

\emph{System without external potentials (V=0).-} When $\lambda >0$ and $V=0$ it was shown in Ref. \cite{Porras} how  
for large $r$ values the leading order approximation to the solution is $R(r) \sim 1/\sqrt{r}$.
When $\lambda < 0$ there are no stationary solutions (this was conjectured in Ref. \cite{Porras}) and will be proven here for later convenience. First we write $A(\boldsymbol{r}) e^{i \varphi(\boldsymbol{r})} = u + i v$, then Eqs. (\ref{gp_dimensionless}) become
\begin{subequations}
\label{RD}
\begin{eqnarray}
\label{sistemaRD:1} -\frac{1}{2}\Delta  u &=&\lambda  u - g(u^2+v^2)u+\gamma (u^2+v^2)^2v , \\
\label{sistemaRD:2} -\frac{1}{2}\Delta  v &=&\lambda  v - g(u^2+v^2)v-\gamma (u^2+v^2)^2u .
\end{eqnarray}
\end{subequations}
Let us now consider Eqs. (\ref{RD}) on a domain $\Omega$ given by a generic ball $B_{\rho}$ with boundary conditions $u=v=0$ on $\partial \Omega$ and $u,v$ to be radially symmetric solutions. Multiplying the first of these equations by  $u$, the second by $v$, integrating in $\Omega$ and applying Green formulae we get 
\begin{multline}
\frac{1}{2} \int  _{B_{\rho}} |\nabla u|^2 +|\nabla v|^2-\pi \rho \left.\left(uu_r+vv_r\right)\right|_{r=\rho}\\ 
= \lambda\int  _{B_{\rho}}  u^2+v^2  -g\int  _{B_{\rho}} (u^2+v^2)^2. \label{green:+:unbounded}
\end{multline} 
Since $R \geq 0$ and $R \to 0$ at infinity, we can always take a sequence  $\rho_k\to\infty$ such that $R_r(\rho_k)\leq 0. $ However  $(R^2)_r=2(uu_r+vv_r),$ therefore the l.h.s. of Eq. (\ref{green:+:unbounded}) is positive for any $B_{\rho_k}.$ As a conclusion, if $g\geq 0$ there are no radially symmetric solutions.

When $\lambda <0,$  $g< 0$ and $V=0$ there are not  radially symmetric solutions with  amplitude in $L^6(\mathbb R^2).$  By comparison results on the Eq. \eqref{r-R:3}, it is not difficult to prove that, when $r\gg 1,$ $R(r)\geq C\exp(\sqrt{-\lambda}\, r)$ for some $C\in \mathbb{R},$  which is unbounded when $r\to \infty$ provided $\lambda<0$.

In summary, when $\lambda <0$ there are neither radially symmetric solutions at all for $g \geq 0$ nor solutions satisfying the finiteness requirement for $Q$ (mathematically $A \in L^6(\mathbb R^2)$ norm) for $g<0$. Thus, we can ensure that there are no physically relevant stationary solutions when $\lambda < 0$.

\emph{Nonexistence of dissipative solitons in systems with rigid walls.-} Now we move to the main point of this paper, which is showing that dissipative solitons of Eq. (\ref{gp_dimensionless}) cannot be trapped by strongly confining potentials, or equivalently proving that stationary solutions of Eq. (\ref{mod-pha}) do not exist for those potentials. 

 To do so we first consider the case of an infinite potential barrier placed on the boundaries of a finite region $\Omega$. Energy conservation then imposes the condition that $u(\boldsymbol{r}) = 0$ for $\boldsymbol{r} \in \partial \Omega$. 
 
We start from Eqs. (\ref{RD}), multiply Eq. (\ref{sistemaRD:1}) by $v$,  Eq. (\ref{sistemaRD:2})  by $u$, and integrate over $\Omega$ to get
\begin{subequations}
\begin{eqnarray}\label{p1}
\frac{1}{2}\int  _{  \Omega  } \nabla u\cdot \nabla v  & = & \lambda\int  _{  \Omega  }  uv \nonumber  -g\int  _{  \Omega  } (u^2+v^2) uv \\   & & 
 + \gamma\int  _{  \Omega  } (u^2+v^2)^2 v^2, \\
\frac{1}{2}\int  _{  \Omega  } \nabla u\cdot \nabla v\nonumber & = & \lambda\int  _{  \Omega  }  uv -g\int  _{  \Omega  } (u^2+v^2) uv  \\ 
& & 
 -\gamma\int  _{  \Omega  } (u^2+v^2)^2 u^2. \label{pp2}
\end{eqnarray}
\end{subequations}
 The difference of  Eqs. (\ref{p1}) and (\ref{pp2}) leads to 
\begin{equation}
\gamma\int  _{  \Omega  } (u^2+v^2)^3 =0, \label{green:-}
\end{equation} 
from where we conclude that either $\gamma =0$ or the only solution is the trivial one $u=v=0$. Thus dissipative solitons cannot exist when confined to a finite region.

We can provide some extra results valid  for more general geometries of the domain $\Omega$. Let us consider again Eqs.  \eqref{RD}. Multiplying Eq. \eqref{sistemaRD:1}  by
 $u$, Eq. \eqref{sistemaRD:2}  by $v$, integrating in $\Omega$ and applying Green formulae we get 
\begin{equation}
\frac{1}{2}\int  _{  \Omega  } |\nabla u|^2  =  \lambda\int  _{  \Omega  }  u^2  -g\int  _{  \Omega  } (u^2+v^2) u^2
+ \gamma\int  _{  \Omega  } (u^2+v^2)^2 uv,\nonumber
\end{equation}
\begin{equation}
\frac{1}{2}\int  _{  \Omega  } |\nabla v|^2  =  \lambda\int  _{  \Omega  }  v^2  -g\int  _{  \Omega  } (u^2+v^2) v^2
-\gamma\int  _{  \Omega  } (u^2+v^2)^2 uv.\nonumber
\end{equation} 
% \begin{multline}
% \frac{1}{2}\int  _{  \Omega  } |\nabla u|^2  =  \lambda\int  _{  \Omega  }  u^2 \\ -g\int  _{  \Omega  } (u^2+v^2) u^2
% + \gamma\int  _{  \Omega  } (u^2+v^2)^2 uv,
% \end{multline}
% \begin{multline}
% \frac{1}{2}\int  _{  \Omega  } |\nabla v|^2  =  \lambda\int  _{  \Omega  }  v^2 \\ -g\int  _{  \Omega  } (u^2+v^2) v^2
% -\gamma\int  _{  \Omega  } (u^2+v^2)^2 uv.
% \end{multline} 
Adding both equations we obtain
\begin{equation}
\frac{1}{2} \int  _{  \Omega  } |\nabla u|^2 +|\nabla v|^2= \lambda\int  _{  \Omega  }  u^2+v^2 -g\int  _{  \Omega  } (u^2+v^2)^2. \label{green:+}
\end{equation} 
Poincare's inequality on bounded domains $\Omega$ says that if $\delta_1$ is the first eigenvalue of
 the Laplacian operator on $\Omega$ with Dirichlet homogeneous boundary conditions, then 
\begin{equation}
\delta_1=\inf _{\phi\in H^1_0(\Omega)}{\int_{\Omega}|\nabla
  \phi|^2\over \int_{\Omega}|\phi|^2}.
\end{equation} 
This inequality also holds for unbounded domains  $\Omega$, which are bounded along one of the directions, which allows
 us to extend our results to systems with strong confinement along a single direction. In any of those situations we will have 
\begin{multline}
\frac{\delta_1}{2}\int  _{  \Omega  }  \left( u^2+v^2 \right)
\leq \frac{1}{2}\int  _{  \Omega  } |\nabla u|^2 +|\nabla v|^2 \\ = \lambda\int  _{  \Omega  }  \left( u^2+v^2 \right)-g\int  _{  \Omega  } (u^2+v^2)^2,
\end{multline} 
from where we find that 
\begin{equation}
\left(\frac{\delta_1}{2}-\lambda \right) \int  _{  \Omega  }  \left(u^2+v^2\right)  \leq  -g\int  _{  \Omega  } (u^2+v^2)^2,
\end{equation} 
so that we get that when $g>0$ the negativity of the second term implies that $\lambda > \delta_1/2$, thus if $\lambda <\delta_1/2$ the only 
solution is $u= v= 0$. Moreover  $\delta_1(\Omega )\to 0$ when $|\Omega |\to \infty$, which matches our previous results on radially symmetric stationary solutions.
However, here the result is proven rigorously for any type of solution (not only symmetric ones) in hard-wall type potentials corresponding to semi-infinite regions.

\emph{Systems with ``strong" confinement .-} We have already shown that dissipative solitons of Eq. (\ref{gp_dimensionless}) cannot be confined to finite spatial regions by hard-wall type potentials.
 In what follows we will study the effect of nonsingular potentials.To do so we will first  start by considering unbounded asymptotically increasing potentials i.e. potentials with $V \rightarrow \infty$
  as $r\rightarrow \infty$, such as those of the form $0< V(r) \sim r^p$ with $p>0$. This is the case of applications to Bose-Einstein condensates in magnetic traps, where usually $V(r) \sim r^2$. These potentials provide 
 a strong confinement and in the linear ($g=0, \gamma=0$) case have an infinite number of bound states.

Let us then  consider the version of Eq. \eqref{green:+:unbounded}
including the potential term. The same type of arguments used for the case $V=0, \lambda <0$ can be used to prove that 
if $g\geq 0$
\begin{equation}\label{fraca}
\lambda \geq   {\int_{B_{\rho_k}} V R^2\over \int_{B_{\rho}}R^2}.
\end{equation}
As stated before,  stationary solutions must have necessarily  $\int_{B_{\rho}}R^2 \to \infty$ when $\rho \to \infty$. For asymptotically increasing potentials there is some $r = r_*$ such that for all $r > r_*$, $V(r_*) > \lambda$ and then both the numerator and denominator of Eq. (\ref{fraca}) diverge  and using L'Hopital rule
\begin{equation}
\lim_{\rho\to\infty} {\int_{B_{\rho}} V R^2\over \int_{B_{\rho}}R^2}=\lim_{\rho\to\infty}{ V R^2\over R^2},
\end{equation}
therefore $\lambda\geq \lim_{\rho\to\infty} V(\rho)= \infty$ and as a conclusion, there are no radially symmetric solutions.

When $g< 0,$ there are no radially symmetric solutions with amplitude in $L^6(\mathbb R^2).$   The reason is again that, when $r\gg 1,$ $R(r)\geq C\exp(\sqrt{-\lambda}\, r)$ for some constant $C,$  which 
is unbounded when $r\to\infty.$ This excludes the possibility of existence of physically meaningful stationary solutions for both signs of the nonlinear coefficient $g$ 
and implies that dissipative solitons of Eq. (\ref{gp_dimensionless}) cannot exist when an asymptotically increasing potential is present.

\emph{Systems with weak confinement.- }  We will now see how dissipative solitons survive when a bounded potential (weak confinement) is present,  i.e.  $V(r) \rightarrow V_\infty$ when $r\rightarrow \infty$. In this situation, arguing as in the previous case, 
\begin{equation}
\lambda \geq   \inf _{\phi}{\int_{B_\rho} V
  \phi^2\over \int_{B_\rho}\phi ^2}\geq \liminf_{\rho\to\infty} V= V_\infty,
\end{equation}
thus implying the possibility of existence of solitary waves only for values of the eigenvalue $\lambda$ satisfying $\lambda\geq V_\infty$.
In fact, the asymptotic analysis of the problem for $r \gg 1$ in this case is the same presented in Refs. \cite{Porras,Ou} and we can expect the existence of solutions which 
will be close to those of the case $V=0$ but with local deformations in the region where $V$ has appreciable values.

\emph{Physical implications and soliton mobility.-} These results have practical implications  for the experimental generation of dissipative solitons in Bose-Einstein condensates with appreciable losses due to three-body collisions. In realistic condensates, due to the condition $N<\infty$ we can only construct dissipative structures close to the stationary solutions but with a finite lifetime \cite{Ou}.
In fact, numerical simulations of blow up phenomena with high order losses and finite number of particles show the formation of multi-ring structures whose shape resemble that of dissipative solitons \cite{Ou,cBEC1}. However, 
our results imply that the magnetic confining potentials of the form $V(r) \sim r^2$ are not appropriate to  confine the atomic cloud if these dissipative solitons are to be obtained. In this sense it would be more convenient to use a laser beam to generate an optical dipole type trap, of the form $V(r) \sim e^{-r^2}$ thus being compatible with the existence of stationary structures. 

Could a localized potential  be used to control these solitons? The problem of soliton mobility is an interesting one in many different contexts \cite{mob1}. In our case an optical dipole trap (or any other ``localized potential") cannot probably be used to move the soliton. The reason is that
 these dissipative solitons arise as a global equilibrium between an inward flux of particles due to the nontrivial phase structure and the central peak at which particles are dissipated \cite{Ou}. So, even if a local potentials could be used to move the central peak,  
 the refilling mechanism would lead to a new soliton at the center \cite{Dubietis2} while the moving part would experience a fast decay due to losses. 
The only possibility for moving this soliton would be to use a bounded potential (to ensure existence) having a very slow decay ensuring a global, although small, action on the dissipative soliton (i.e. on the same scale of  the soliton). An example could be a potential of  Yukawa type  $V(r) \sim \left[1-\exp(-r)\right]/r$.

\emph{Conclusions.-} In summary, we have studied a class dissipative solitons in models which have received a lot of attention in the last years in different physical contexts \cite{Dubietis,Porras,O-waves,Ou,Sulem,cBEC0,cBEC1,cBEC3,cBEC4,cBEC5,cBEC6,cBEC3b,Zakharov,dissipation}. Our main result is that contrary to the intuition and the observed typical behavior of solitons these dissipative solitons cannot be trapped by using strongly confining  potentials. This fact represents a notable exception to the typical behavior of solitons in the context of systems ruled by nonlinear Schr\"odinger equations. 
We have also discussed some physical implications of our findings and made some conjectures on the mobility of these solitons on the basis of our findings.

This work has been partially supported by grants BFM2003-02832 and MTM2006-08262 (Ministerio de Educaci\'on y Ciencia, Spain)
and PAI-05-001 (Consejer\'{\i}a de Educaci\'on y Ciencia de la Junta de Comunidades de Castilla-La Mancha).

\end{document}